\documentclass[amsmath,amssymb,aps,pra,reprint,superscriptaddress]{revtex4-2}

\usepackage{graphicx}
\usepackage{dcolumn}
\usepackage{bm}
\usepackage{booktabs}
\usepackage{array}
\usepackage{float}
\usepackage{multirow}
\usepackage{braket}
\usepackage{gensymb}
\usepackage[colorlinks=true,linkcolor=blue,citecolor=blue,urlcolor=blue]{hyperref}
\begin{document}

\title{Polarization-entangled photon pairs generation from a single lithium niobate waveguide with single poling period}

\author{Xinyue Zhang}
\affiliation{State Key Laboratory of Extreme Photonics and Instrumentation, College of Optical Science and Engineering, Zhejiang University, Hangzhou 310027, China}
\author{Sihui Pei}
\affiliation{State Key Laboratory of Extreme Photonics and Instrumentation, College of Optical Science and Engineering, Zhejiang University, Hangzhou 310027, China}
\author{Ni Yao}
\affiliation{Research Center for Frontier Fundamental Studies, Zhejiang Lab, Hangzhou 311100, China}
\author{Shuhao Wang}
\affiliation{State Key Laboratory of Extreme Photonics and Instrumentation, College of Optical Science and Engineering, Zhejiang University, Hangzhou 310027, China}
\author{J. Q. You}
\affiliation{Zhejiang Key Laboratory of Micro-Nano Quantum Chips and Quantum Control, School of Physics, Zhejiang University, Hangzhou 310027, China}
\author{Limin Tong}
\affiliation{State Key Laboratory of Extreme Photonics and Instrumentation, College of Optical Science and Engineering, Zhejiang University, Hangzhou 310027, China}
\affiliation{Jiaxing Key Laboratory of Photonic Sensing $\rm{\&}$ Intelligent Imaging, Intelligent Optics $\rm{\&}$ Photonics Research Center, Jiaxing Research Institute Zhejiang University, Jiaxing 314000, China}
\author{Wei Fang}
\email{wfang08@zju.edu.cn}
\affiliation{State Key Laboratory of Extreme Photonics and Instrumentation, College of Optical Science and Engineering, Zhejiang University, Hangzhou 310027, China}
\affiliation{Jiaxing Key Laboratory of Photonic Sensing $\rm{\&}$ Intelligent Imaging, Intelligent Optics $\rm{\&}$ Photonics Research Center, Jiaxing Research Institute Zhejiang University, Jiaxing 314000, China}

\date{\today}

\begin{abstract}
	Polarization-entangled photon pairs are essential sources for photonic quantum information processing. However, generating entangled photon pairs with large detuning via spontaneous parametric down-conversion (SPDC) often requires complex configurations to compensate for phase matching. Here, we propose a simple and efficient scheme to generate polarization-entangled photon pairs based on type-0 SPDC in a thin-film lithium niobate waveguide with a single poling period. By utilizing the strong dispersion engineering capabilities of thin-film waveguides, we can achieve both degenerate and highly detuned entangled photon pairs. Furthermore, we demonstrate on-chip temporal compensation using an integrated waveguide structure. Our approach offers a compact and scalable solution for integrated quantum photonic circuits.
\end{abstract}

\maketitle

\section{Introduction}
	Quantum entanglement is a fascinating quantum state that makes quantum information technology possible and serves as a crucial tool for exploring quantum theory \cite{horodecki2009quantum,vedral2014quantum,chen2021quantum}. Polarization-entangled photon pairs are one of the most common entangled sources because of the relatively easy generation and manipulation \cite{edamatsu2007entangled}. There are primarily three methods for generating polarization-entangled photon pairs: spontaneous parametric down-conversion (SPDC) in second-order $\chi^2$ nonlinear materials \cite{anwar2021entangled}, four-wave mixing in third-order  $\chi^3$ nonlinear waveguides \cite{li2005optical,zhu2012direct,matsuda2016generation}, and cascade transitions of biexcitons in semiconductor quantum dots \cite{garcia2021semiconductor,schimpf2021quantum}. SPDC is particularly noteworthy due to its high conversion efficiency and ability to operate at room temperature \cite{kitaeva2005spontaneous,couteau2018spontaneous}. In nonlinear crystals, dispersion and birefringence typically cause the down-converted photon pairs to emerge non-collinearly, with entanglement only occurring in the overlap of signal and idler cones \cite{kwiat1995new,kwiat1999ultrabright}. The introduction of periodic poling compensates for momentum mismatch, enabling collinear output of down-converted photon pairs \cite{kuklewicz2004high,gao2024polarization}. Furthermore, the integration of waveguide structures, which confine the light field within a small space, significantly enhances nonlinear effects \cite{zhu2021integrated}. However, photons generated via SPDC in a single decay channel are not inherently entangled in the polarization degree of freedom. To obtain a maximum Bell state, two decay paths with orthogonal polarization states from type-I/0 ($\ket{H_s}\ket{H_i}$ and $\ket{V_s}\ket{V_i}$) or type-II ($\ket{H_s}\ket{V_i}$ and $\ket{V_s}\ket{H_i}$) must be utilized, along with indistinguishable spatial, spectral, and temporal features.In practice, polarization-entangled degenerate (or non-degenerate with small detuning) photon pairs can be generated from a single periodically poled nonlinear crystal waveguide based on a type-II scheme \cite{duan2020generation}. Non-degenerate pairs with large detuning can be realized using either two crystals \cite{pelton2004bright,ljunggren2006theory,hubel2007high,steinlechner2012high,scheidl2014crossed}, a single crystal waveguide with two poling periods \cite{suhara2009quasi,thomas2010novel,herrmann2013post,sun2019compact,shukla2020generation,shukla2021polarization}, or through double pass techniques in the same waveguide using interferometric setups such as Michelson \cite{martin2010polarization}, Mach-Zehnder \cite{suhara2007generation,martin2009integrated,zhong2010high,kaiser2013versatile}, and Sagnac structures \cite{lim2008stable,arahira2011generation,vergyris2017fully,meyer2018high}.
	
	In this work, utilizing the strong dispersion control capabilities of thin-film lithium niobate (LN) waveguides, we propose generating polarization-entangled photon pairs based on a type-0 scheme in a single waveguide with a single poling period. By adjusting the polarization state of the pump light, a maximum Bell state can be achieved. The photon pairs can be degenerate or non-degenerate with large detuning. Additionally, temporal compensation can be achieved on the same chip by integrating an extra waveguide segment.

\section{Thin-film lithium niobate waveguide configuration} 
	LN is an excellent nonlinear material with high second-order nonlinear susceptibility coefficients, which can be described by two-dimensional tensor $d_{il}$: 
	\begin{equation}\label{eq1}
	d_{il} = \begin{bmatrix} 
				0 & 0 & 0 & 0 & d_{31} & -d_{22} \\ 
				-d_{22} & d_{22} & 0 & d_{31} & 0 & 0 \\
				d_{31} & d_{31} & d_{33} & 0 & 0 & 0
			 \end{bmatrix}.
	\end{equation}
	\begin{table*}[t]
		\centering
		\caption{\label{table1}Possible configurations to achieve SPDC in a PPLN waveguide.}
		\begin{ruledtabular}
			\begin{tabular}{c c c c c c c c}
				\multirow{2}{*}{Configuration \#} &
				\multirow{2}{*}{Type} &
				\multirow{2}{*}{Crystal cut} &
				\multirow{2}{*}{Waveguide orientation} &
				\multicolumn{3}{c}{Polarization state of} &
				\multirow{2}{*}{$d_{ij}$} \\
				\cmidrule(r){5-7}
				& & & & pump & signal & idler & \\
				\midrule
				
				\multirow{2}{*}{1} & \multirow{8}{*}{II} & \multirow{2}{*}{z} & \multirow{2}{*}{x} & TE & TE & TM & \multirow{2}{*}{$d_{31}$}\\
				& & & & TE & TM & TE & \\
				\cmidrule(r){1-1} \cmidrule(r){3-8}
				
				\multirow{2}{*}{2} & & \multirow{2}{*}{z} & \multirow{2}{*}{y} & TE & TE & TM & \multirow{2}{*}{$d_{31}$} \\
				& & & & TE & TM & TE & \\
				\cmidrule(r){1-1} \cmidrule(r){3-8}
				
				\multirow{2}{*}{3} & & \multirow{2}{*}{x} & \multirow{2}{*}{y} & TM & TE & TM & \multirow{2}{*}{$d_{31}$} \\
				& & & & TM & TM & TE & \\
				\cmidrule(r){1-1} \cmidrule(r){3-8}
				
				\multirow{2}{*}{4} & & \multirow{2}{*}{y} & \multirow{2}{*}{x} & TM & TE & TM & \multirow{2}{*}{$d_{31}$} \\
				& & & & TM & TM & TE & \\
				\cmidrule(r){1-8}
				
				\multirow{2}{*}{5} & \multirow{8}{*}{0+I} & \multirow{2}{*}{x} & \multirow{2}{*}{y} & TE & TE & TE & $d_{33}$ \\
				& & & & TE & TM & TM & $d_{31}$ \\
				\cmidrule(r){1-1} \cmidrule(r){3-8}
				
				\multirow{2}{*}{6} & & \multirow{2}{*}{y} & \multirow{2}{*}{x} & TE & TE & TE & $d_{33}$ \\
				& & & & TE & TM & TM & $d_{31}$ \\
				\cmidrule(r){1-1} \cmidrule(r){3-8}
				
				\multirow{2}{*}{7} & & \multirow{2}{*}{z} & \multirow{2}{*}{x} & TM & TE & TE & $d_{31}$ \\
				& & & & TM & TM & TM & $d_{33}$ \\
				\cmidrule(r){1-1} \cmidrule(r){3-8}
				
				\multirow{2}{*}{8} & & \multirow{2}{*}{z} & \multirow{2}{*}{y} & TM & TE & TE & $d_{31}$ \\
				& & & & TM & TM & TM & $d_{33}$ \\
				\cmidrule(r){1-8}	
				
				\multirow{2}{*}{9} & \multirow{4}{*}{0} & \multirow{2}{*}{y} & \multirow{2}{*}{x} & TE & TE & TE & $d_{33}$ \\
				& & & & TM & TM & TM & $d_{22}$ \\
				\cmidrule(r){1-1} \cmidrule(r){3-8}
				
				\multirow{2}{*}{10} & & \multirow{2}{*}{z} & \multirow{2}{*}{x} & TE & TE & TE & $d_{22}$ \\
				& & & & TM & TM & TM & $d_{33}$ \\		
			\end{tabular}
		\end{ruledtabular}
	\end{table*}
	
	\noindent Based on this tensor, all the possible configurations for SPDC processes in a straight periodically poled waveguide are listed in Table 1. The type-II scheme can down-convert a pump photon with a specific polarization into a pair of orthogonally polarized photons. However, due to the different dispersion properties of the two polarization modes in the waveguide, achieving maximal entanglement is challenging. To ensure high concurrence, the wavelengths of the signal and idler photons should not differ significantly in a waveguide with a single poling period \cite{duan2020generation}. Additionally, the significant $d_{33}$ coefficient of the LN crystal cannot be utilized. Using a type-I configuration alone cannot generate polarization-entangled photons in a single waveguide; it requires the assistance of a type-0 configuration to produce photon pairs of the other polarization. Nevertheless, both configurations use the same polarization for the pump light, they encounter challenges similar to those of type-II. Here we focus on the type-0 scheme, where TE/TM polarized pump photons generate TE/TM polarized signal and idler photons, respectively. This setup enables the generation of a maximum Bell state by adjusting the polarization of the pump light, with fewer wavelength constraints.
	
	To generate an entangled state like $1/\sqrt2(\ket{H_s}\ket{H_i}+\ket{V_s}\ket{V_i}$ in a single waveguide with a single periodic poling, it is necessary to satisfy the phase matching conditions for both down-conversion processes simultaneously. Additionally, the signal and idler photons produced in both processes must be indistinguishable in terms of wavelength and timing. Fortunately, the strong dispersion control capabilities of thin-film LN waveguides make this feasible.
	
	\begin{figure}[b]
		\includegraphics{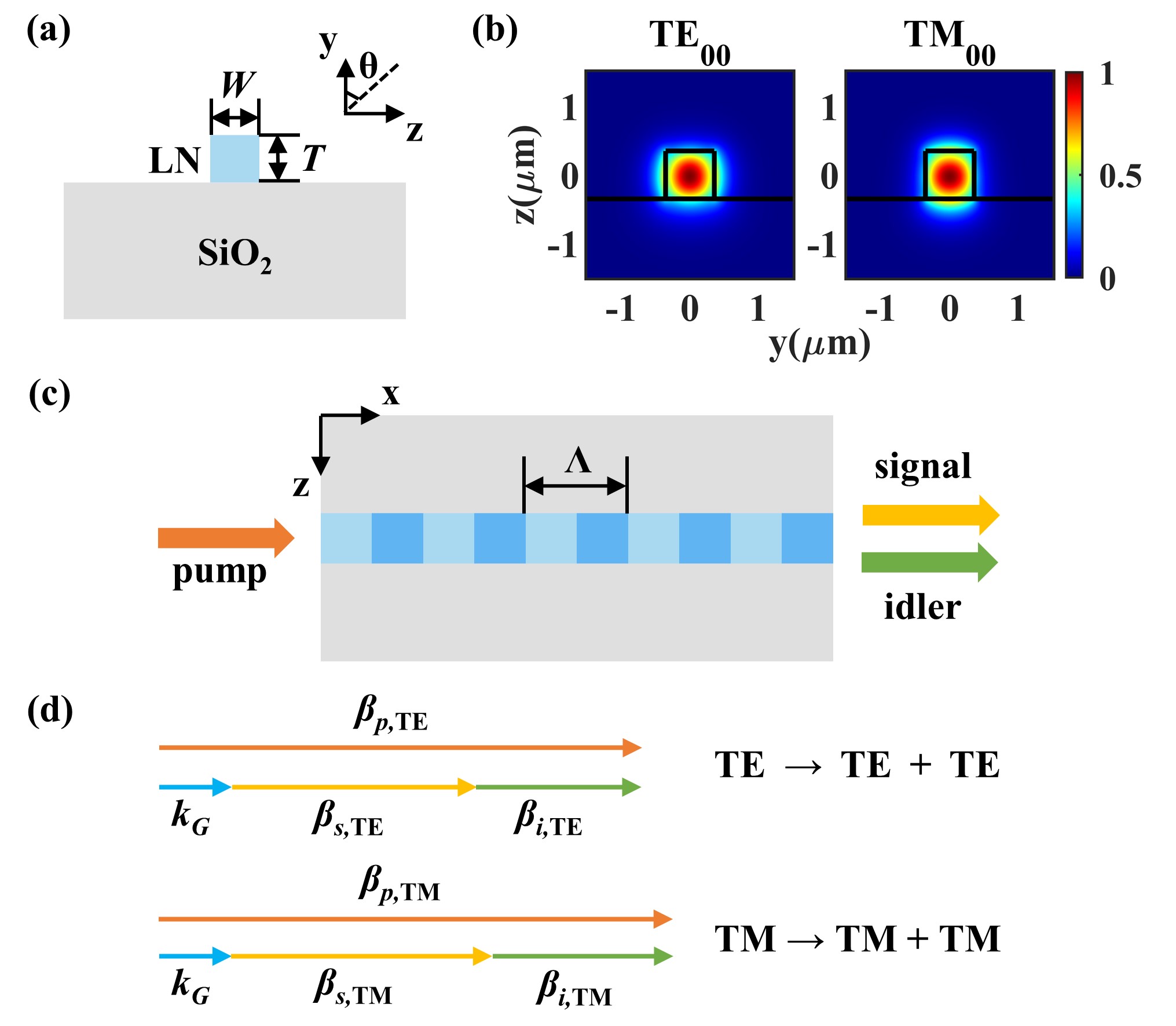}
		\caption{\label{fig1} (a) Cross-section of a y-cut thin-film PPLN waveguide. (b) Mode profiles of $\mathrm{TE_{00}}$ and $\mathrm{TM_{00}}$ at wavelength of 1550 nm. (c) SPDC process in a thin-film PPLN waveguide with a single poling period. (d) QPM conditions for two type-0 SPDC processes with different polarizations.}
	\end{figure}
	
	\begin{figure*}[t]
		\includegraphics{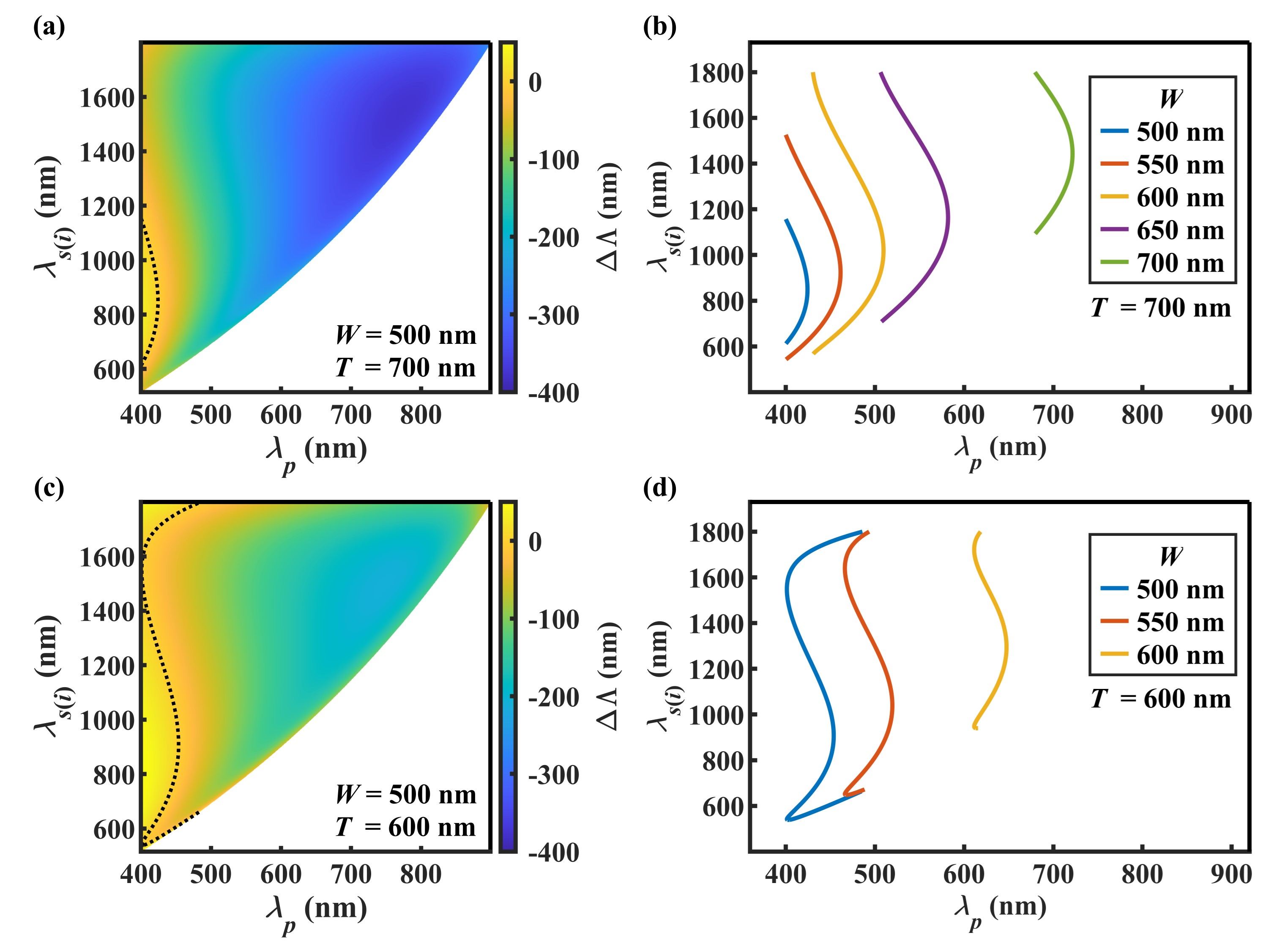}
		\caption{\label{fig2} (a) and (c), color-coded maps of poling period difference ($\Delta\Lambda$) for TE and TM polarizations as functions of pump and signal (idler) wavelengths in two waveguide configurations. Dotted lines mark regions satisfying the single poling period condition ($\Delta\Lambda = 0$). Corresponding single poling period conditions for different waveguide parameters are shown in (b) and (d).}
	\end{figure*}
	
	As an example, Fig.~\ref{fig1}(a) shows the cross-section of a y-cut x-orientation thin-film periodically poled lithium niobate (PPLN) waveguide (configuration \#9) on a $\mathrm{SiO_2}$ substrate, which has a width $W$ and height $T$. Here we only consider the fundamental $\mathrm{TE_{00}}$ and $\mathrm{TM_{00}}$ modes (as shown in Fig.~\ref{fig1}(b)), corresponding to the $H$ and $V$ polarized output photons, respectively. For a PPLN waveguide with single poling period of $\Lambda$, as illustrated in Fig.~\ref{fig1}(c), efficient SPDC occurs for both modes when the conservation laws are satisfied. The energy conservation law can be expressed as: 
	\begin{equation}\label{eq2}
		\omega_{s} + \omega_{i} = \omega_{p},
	\end{equation}
	where $\omega_{s}$, $\omega_{i}$ and $\omega_{p}$ represent the angular frequency of signal, idler and pump photons, respectively. The quasi-phase matching (QPM) condition requires: 
	\begin{equation}\label{eq3}
	\beta_{p,\mathrm{TE}}-\beta_{s,\mathrm{TE}}-\beta_{i,\mathrm{TE}}-k_{G} = \beta_{p,\mathrm{TM}}-\beta_{s,\mathrm{TM}}-\beta_{i,\mathrm{TM}}-k_{G} = 0,
	\end{equation}	
	where $k_{G} = 2m\pi/\Lambda$, and $\beta_{j,q}(j=p,s,i;q=\mathrm{TE,TM})$ represents the propagation constants of the pump, signal and idler light of TE (TM) modes, respectively, as indicated in Fig.~\ref{fig1}(d). Due to the different dispersion relations of the polarization modes, the poling periods required to satisfy the QPM conditions for type-0 SPDC for signal and idler photons of various wavelength are different in general. Therefore, the requirement of a single poling period imposes a restriction on the wavelengths of the pump and down-converted photons, which in turn determines the bandwidth of the generated down-converted photons.
	
	Figure~\ref{fig2}(a) depicts the wavelengths of signal (idler) photons with identical wavelengths for both polarization states, generated at a specific pump wavelength in a waveguide with $T$ = 700 nm and $W$ = 500 nm. The color indicates the poling period difference ($\Delta\Lambda$) for the two polarization states that satisfy the QPM conditions. From this figure, the curve that satisfies single poling period condition ($\Delta\Lambda = 0$) can be directly obtained. The large refractive index contrast allows the thin-film LN waveguide to achieve strong dispersion control by adjusting the waveguide's thickness and width, thereby providing strong tuning capabilities for wavelengths of down-converted photon pairs. Figure~\ref{fig2}(b) illustrates the distribution of signal and idler photon wavelengths in a single-poling period waveguide with a thickness of $T$ = 700 nm, for various waveguide widths. Therefore, specific wavelengths of entangled photon pairs can be achieved by adjusting the waveguide's thickness, width, and the pump light wavelength parameters. Additionally, under certain waveguide parameter conditions, as shown in Fig.~\ref{fig2}(c) and (d), more complex curves provide the possibility of generating two sets of entangled photon pairs with different wavelengths from the same waveguide. 
	
\section{Pump light scheme}
	Section 2 discusses the waveguide structure parameters required to meet the phase matching conditions for generating orthogonally-polarized photon pairs in a single-period PPLN waveguide. In this section, we will discuss the setup of the pump light to ensure that the photon pairs are polarization entangled. For simplicity, we consider the pump field as monochromatic, continuous-wave light. Take configuration \#9 for example, as shown in Fig.~\ref{fig1}(a), the two-photon polarization-entangled state to be expressed as:
	\begin{widetext}
	\begin{equation}\label{eq4}
	\begin{split}
		\vert\Psi\rangle = & M_{\mathrm{TE}} d_{\mathrm{TE}} \int\mathrm{d}\upsilon \int_{-L}^{0}\mathrm{d}x e^{i(\beta_{p,\mathrm{TE}}-\beta_{s,\mathrm{TE}}-\beta_{i,\mathrm{TE}}-k_{G})x}\hat{a}^{\dagger}_{s,\mathrm{TE}}(\omega_{s0}+\upsilon) \hat{a}^{\dagger}_{i,\mathrm{TE}}(\omega_{i0}-\upsilon) \ket{vac} \\
		 & + M_{\mathrm{TM}} d_{\mathrm{TM}} \int\mathrm{d}\upsilon \int_{-L}^{0}\mathrm{d}x e^{i(\beta_{p,\mathrm{TM}}-\beta_{s,\mathrm{TM}}-\beta_{i,\mathrm{TM}}-k_{G})x}\hat{a}^{\dagger}_{s,\mathrm{TM}}(\omega_{s0}+\upsilon) \hat{a}^{\dagger}_{i,\mathrm{TM}}(\omega_{i0}-\upsilon) \ket{vac},
	\end{split}
	\end{equation}	
	\end{widetext}
	where coefficients
	\begin{equation}\label{eq5}
		M_q = i\sqrt{\frac{P_q\omega_{s0}\omega_{i0}\xi^2_q}{2\varepsilon_0c^3n_{p,q}n_{s,q}n_{i,q}S_{eff,q}}}(q=\mathrm{TE,TM}),
	\end{equation}	
	the first-order nonlinear coefficient with 0.5 duty cycle is $d_{\mathrm{TE}}=2d_{33}/\pi$, $d_{\mathrm{TM}}=2d_{22}/\pi$, $L$ is the length of waveguide, $\hat{a}^{\dagger}_{s(i),q}$ represents the photon creation operator for the signal (idler) beam with TE (TM) polarization, and $\upsilon$ is the frequency offset of the center frequency $\omega_{j0}(j=s,i)$, assuming $\upsilon\ll\omega_{j0}$. Here the coefficient $M_q$ is slowly varying function of frequency; $P_q$ is the pump power of TE/TM polarization; $n_{j,q}(j=p,s,i)$ represents the effective refractive indices of the LN waveguide for different waves; $\xi_q$ and $S_{eff,q}$ represent the spatial mode overlap factor and effective mode area of three waves in the SPDC process, respectively: 
	\begin{widetext}
	\begin{equation}\label{eq6}
		\xi_q=\frac{\int_{\chi^{(2)}}
		E^*_{p,z(y)}E_{s,z(y)}E_{i,z(y)}\mathrm{d}y\mathrm{d}z} 
		{\left(
			 \left\lvert \int_{\chi^{(2)}}
			 {\lvert \vec{E}_p \rvert}^2 \vec{E}_p \mathrm{d}y\mathrm{d}z
			 \right\rvert
			 \left\lvert \int_{\chi^{(2)}}
			 {\lvert \vec{E}_s \rvert}^2 \vec{E}_s \mathrm{d}y\mathrm{d}z
			 \right\rvert
			 \left\lvert \int_{\chi^{(2)}}
			 {\lvert \vec{E}_i \rvert}^2 \vec{E}_i \mathrm{d}y\mathrm{d}z
			 \right\rvert
		\right)^{\frac{1}{3}}},
	\end{equation}	
	\begin{equation}\label{eq7}
		S_{eff,q}=\left(S_{p,q}S_{s,q}S_{i,q}\right)^{\frac{1}{3}},
		S_{j,q}=\frac{
			\left( \int_{all}
			{\lvert \vec{E}_{j,p} \rvert}^2 \mathrm{d}y\mathrm{d}z
			\right)^3}
			{\left\lvert \int_{\chi^{(2)}}
			{\lvert \vec{E}_{j,p} \rvert}^2 \vec{E}_{j,p} \mathrm{d}y\mathrm{d}z
			\right\rvert^2}
		(j=p,s,i),
	\end{equation}	
	\end{widetext}
	Here $E_{j,p}(j=p,s,i)$ represents the electric field distribution for each mode; $\int_{\chi^{(2)}}$ and $\int_{all}$ correspond to the integration in the nonlinear action region and the whole domain, respectively.  
	
	As the waveguide has a finite length $L$, the integrals $\int_{-L}^{0}\mathrm{d}x$ in Eq.~(\ref{eq4}) result in sinc functions:
	\begin{equation}\label{eq8}
		\begin{split}
		\int_{-L}^{0}\mathrm{d}x e^{i\Delta k_qx} 
		& = Lh(\Delta k_qL) \\
		& = L \exp{(-i\Delta k_qL/2)}\mathrm{sinc}(\Delta k_qL/2), \\
		\end{split}
	\end{equation}	
	where $\Delta k_q=\beta_{p,q}-\beta_{s,q}-\beta_{i,q}-k_{G}$. Owing to different propagation constants, the sinc functions representing the bandwidths of two SPDC processes with different polarizations differ, particularly for non-degenerate photon pairs. Maximum entanglement of the output state is achievable only within the spectral overlap region. This issue can be addressed by employing a narrow bandpass filter to isolate the spectral region where both sinc functions coincide and attain unity. Under this circumstance, the two integrals $\int \mathrm{d}\upsilon$ in Eq.~(\ref{eq4}) yield identical results. Consequently, the condition for achieving a maximally entangled state simplifies to $M_{\mathrm{TE}}d_{\mathrm{TE}}$=$M_{\mathrm{TM}}d_{\mathrm{TM}}$. This condition can be met by adjusting the polarization of the pump beam such that its polarization angle relative to the y-axis satisfies:
	\begin{equation}\label{eq9}
		\tan\theta_m = \sqrt{
			\frac{\xi^2_\mathrm{TM}d^2_\mathrm{TM}}{n_{s,\mathrm{TM}}n_{i,\mathrm{TM}}}
			\biggl/
			\frac{\xi^2_\mathrm{TE}d^2_\mathrm{TE}}{n_{s,\mathrm{TE}}n_{i,\mathrm{TE}}}}.
	\end{equation}	
	And the generation rate of the photon pairs can be estimated by: 
	\begin{equation}\label{eq10}
		\begin{split}
		R & = \langle\Psi\vert\Psi\rangle=2\lvert M_{qs}d_{qs}L \rvert^2 \int\mathrm{d}\upsilon \lvert h(\Delta k_{qs}) \rvert^2 \\
		& = \frac{2\pi L \omega_{s0} \omega_{i0} P_{qs} \xi_{qs}^2 d_{qs}^2}
		{\varepsilon_0 c^3 n_{p,qs} n_{s,qs} n_{i,qs} S_{eff,qs} D_{qs}}. \\
		\end{split}
	\end{equation}	
	Here $qs$ represents the SPDC process which has the narrower bandwidth, and $\Delta k_q$ can be approximated as $\Delta k_q=\upsilon D_q$ by the first-order Taylor expansion, where $D_q=u_{i0,q}^{-1}-u_{s0,q}^{-1}$, $u_{s(i)0,q}$ is the group velocity of corresponding light. The quantum theory analysis is compatible with the classical theory \cite{boyd2008nonlinear,gao2007prediction}. 
	
	\begin{figure}[h]
		\includegraphics{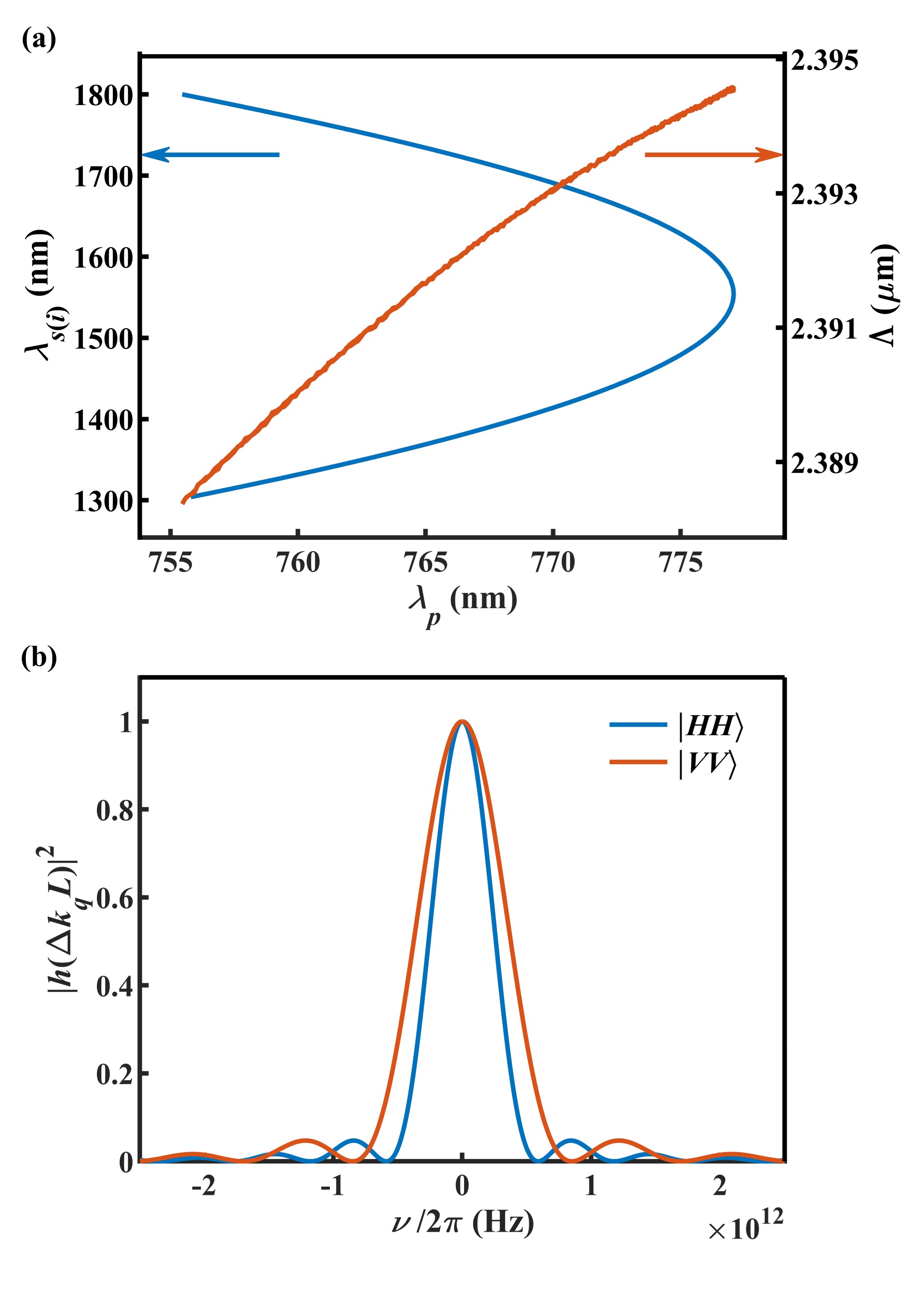}
		\caption{\label{fig3} (a) Signal (idler) wavelength $\lambda_{s(i)}$ as a function of pump wavelength $\lambda_p$ that satisfies QPM condition (blue line), and the corresponding poling periods $\Lambda$ (red line) for a waveguide with a thickness of 700 nm and a width of 712 nm. (b) The calculated bandwidths of the SPDC photon pair, where $\lambda_p$ = 770 nm, $\lambda_s$ = 1413.9 nm, $\lambda_i$ = 1690.8 nm.}
	\end{figure}
	
	As an example, Fig.~\ref{fig3}(a) shows the relationship between the signal (idler) wavelength and the pump wavelength for a PPLN waveguide with a thickness of 700 nm and a width of 712 nm. By adjusting the pump wavelength from 777.08 nm to 755 nm, either degenerate polarization-entangled photon pairs at 1554.16 nm or non-degenerate polarization-entangled photon pairs with a large detuning can be generated. Interestingly, the poling period required to satisfy the QPM condition remains nearly constant (with a difference of less than 10 nm), indicating the potential for broad wavelength tuning of entangled photon pairs within a single PPLN waveguide simply by adjusting the pump wavelength. For applications requiring one photon of the entangled pair to be in the visible range for easy detection and the other in the telecom wavelength range for long-distance propagation, this can be achieved by selecting a waveguide width of 600 nm, as shown in Fig.~\ref{fig2}(b).
	
	The generation rate of the entangled photon pairs can be estimated based on Eq.~(\ref{eq10}). Using the waveguide shown in Fig.~\ref{fig3}(a) as an example, when the pump photon wavelength is 770 nm, the signal and idler photons have wavelengths of 1413.9 nm and 1690.8 nm, respectively. As shown in Fig.~\ref{fig3}(b), the bandwidth of two SPDC processes can be estimated as 0.52 THz and 0.75 THz for $\ket{H_s}\ket{H_i}$ and $\ket{V_s}\ket{V_i}$ states, respectively. Employing a narrow bandpass filter with a 0.52 THz bandwidth and a 1 mW linearly polarized pump at $\theta = 4.276\degree$, maximum entangled photon pairs can be generated with a rate of $R = 6.39\times10^8$ pairs/s. The rate of generation can be increased to $R = 6.04\times10^9$ pairs/s in the nearly degenerate scenario, with $\lambda_p$, $\lambda_s$, and $\lambda_i$ being 777 nm, 1540 nm, and 1568 nm, respectively.
	
	\begin{figure}[h]
		\includegraphics{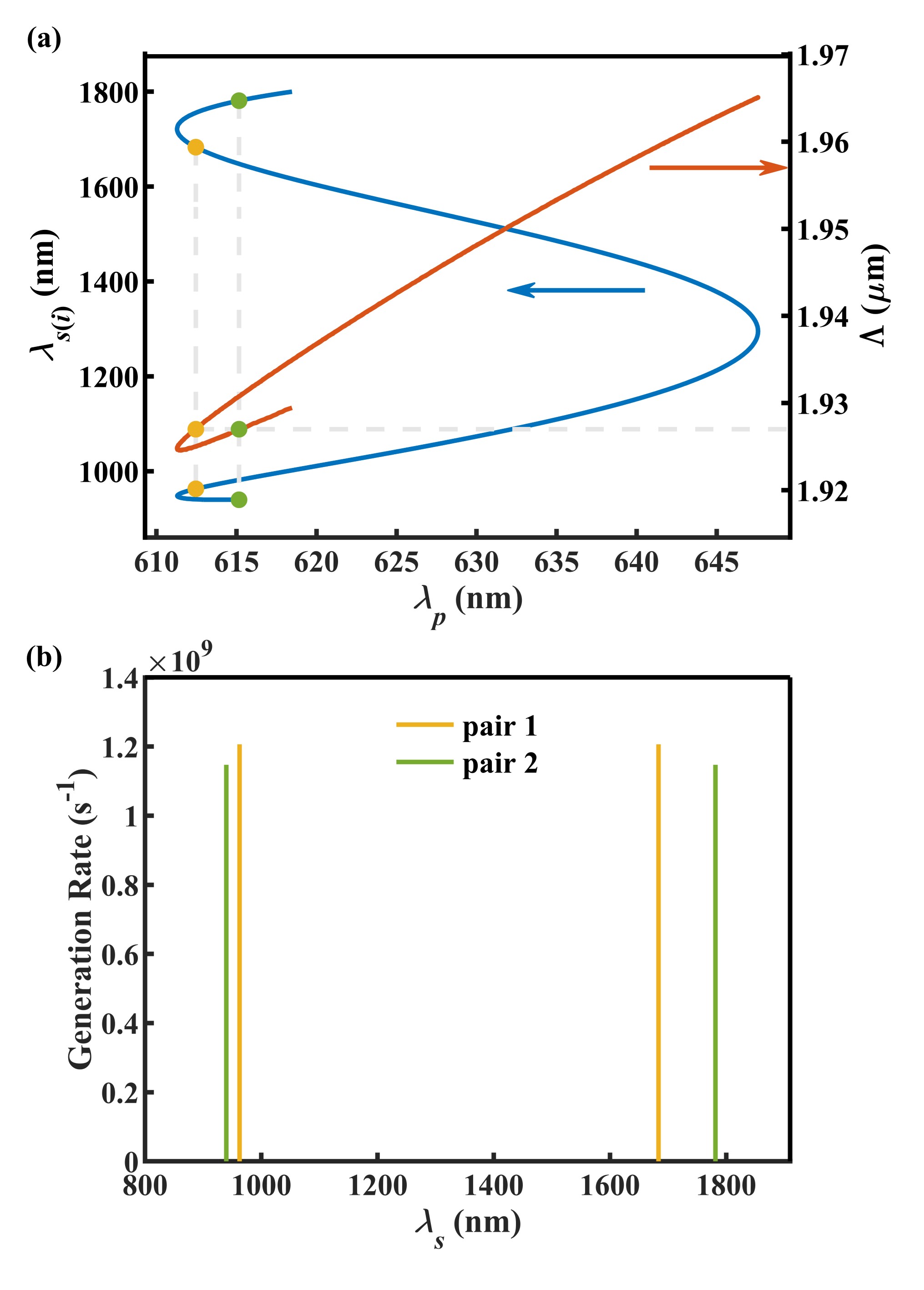}
		\caption{\label{fig4} (a) Signal (idler) wavelength $\lambda_{s(i)}$ as a function of pump wavelength $\lambda_p$ that satisfies QPM condition (blue line), and the corresponding poling periods $\Lambda$ (red line) for a waveguide with the thickness and the width both set as 600 nm. (b) The generation rates of two sets of polarization-entangled photon pairs when the pump power remains 1 mW.}
	\end{figure}
	
	Interestingly, with appropriate waveguide parameters, a single-periodically poled PPLN waveguide can generate two distinct sets of polarization-entangled photon pairs at different wavelengths. As depicted in Fig.~\ref{fig4}(a), when the waveguide’s thickness and width are both set to 600 nm, and the poling period is $1.927~\mathrm{\mu m}$, it is possible to generate two sets of polarization-entangled photon pairs within this waveguide by simply adjusting the pump wavelength and its polarization angle. As illustrated in Fig.~\ref{fig4}(b), when a 1 mW pump light has a wavelength of 612.43 nm and a polarization angle of $4.278\degree$, the first set of polarization-entangled photon pairs is produced with wavelengths of 962.60 nm and 1683.56 nm, at a generation rate of $1.21\times10^9$ pairs/s. Alternatively, when the pump wavelength is adjusted to 615.23 nm and the polarization angle is altered to $4.282\degree$, the second set of polarization-entangled photon pairs emerges with wavelengths of 939.78 nm and 1781.53 nm, at a generation rate of $1.15\times10^9$ pairs/s. As seen in Fig.~\ref{fig2}(d), such scenarios are not uncommon in the type-0 scheme and can cover a wide wavelength range.
	
\section{On-chip temporal compensation}	

	\begin{figure}[b]
		\includegraphics{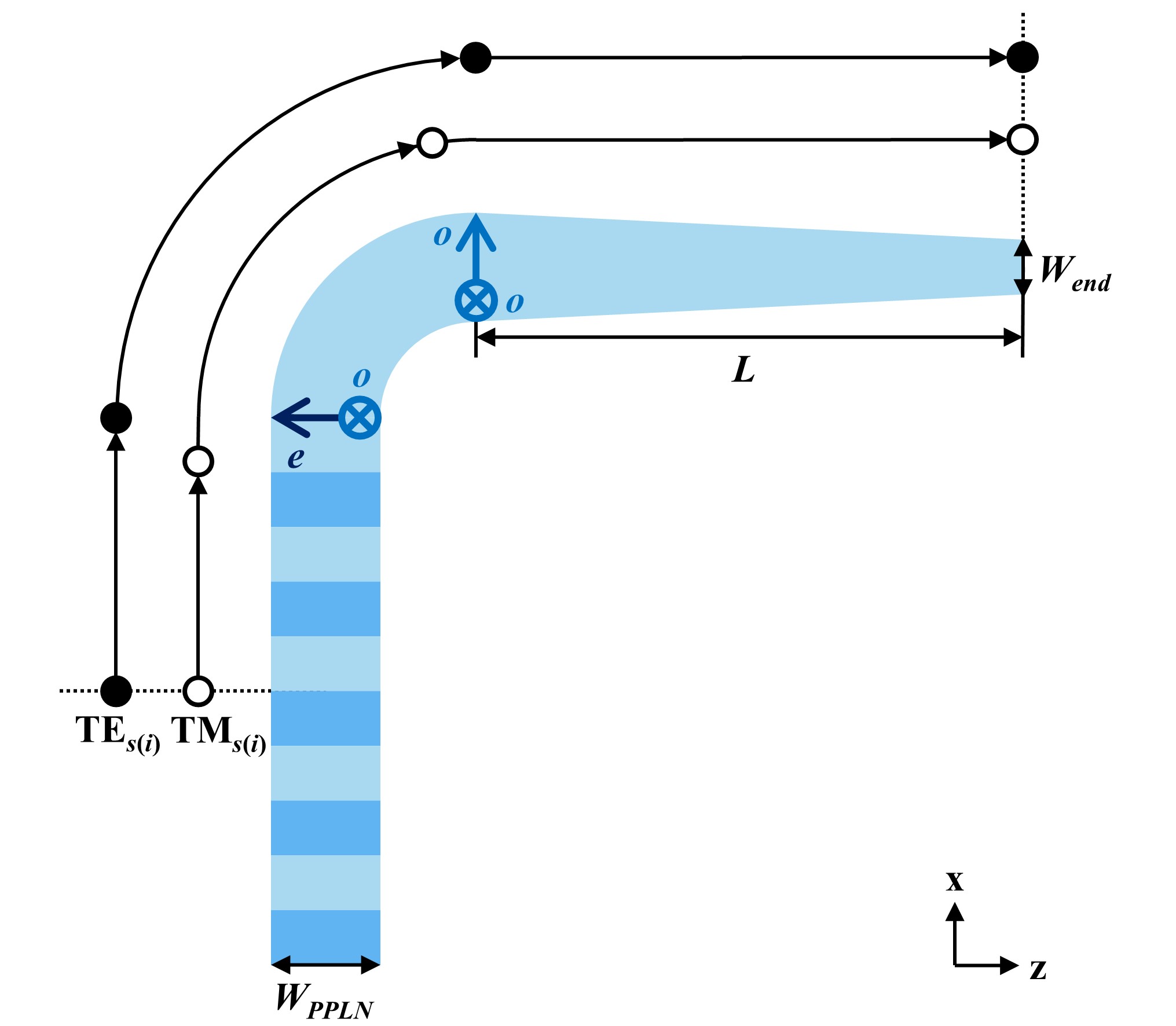}
		\caption{\label{fig5} Schematic diagram of a waveguide structure designed for temporal compensation, consisting of PPLN, bent, and tapered waveguide sections.}
	\end{figure}

	The different effective index between the two polarized modes in the waveguide induces a polarization-dependent phase on the generated SPDC output, thereby degrading the entanglement quality. In general, an additional birefringent crystal is needed for temporal compensation. Fortunately, the planar thin-film LN waveguide structure allows for on-chip temporal compensation through waveguide dispersion control. For example, for the PPLN waveguide discussed in Fig.~\ref{fig3}, the TE and TM modes align with the extraordinary (e) and ordinary (o) refractive indices, respectively. This configuration hinders temporal compensation by adjusting the waveguide width along. However, rotating the waveguide $90\degree$ so that both modes correspond to the ordinary refractive index enables effective compensation via waveguide tuning. 
	
	Figure~\ref{fig5} illustrates the specific waveguide structure. The 1 cm long PPLN waveguide is followed by a waveguide with a $10~\mathrm{\mu m}$ bending radius, which is then connected to a tapered, unpoled waveguide. The tapered waveguide is unable to efficiently generate SPDC photon pairs due to phase mismatch. Thus, it functions solely as a dispersion controller. For simplicity, the tapered waveguide's width linearly decreases, starting from the PPLN waveguide's width of 712 nm. Simulations reveal that when a 3.603 mm long waveguide is tapered to 509.2 nm in width, both the signal (at 1413.9 nm) and idler (at 1690.8 nm) photons with different polarizations generated in the middle of the PPLN waveguide can simultaneously reach the right output port, realizing on-chip temporal compensation.
	
\section{Counter-propagating scheme}
\begin{figure*}[t]
		\includegraphics{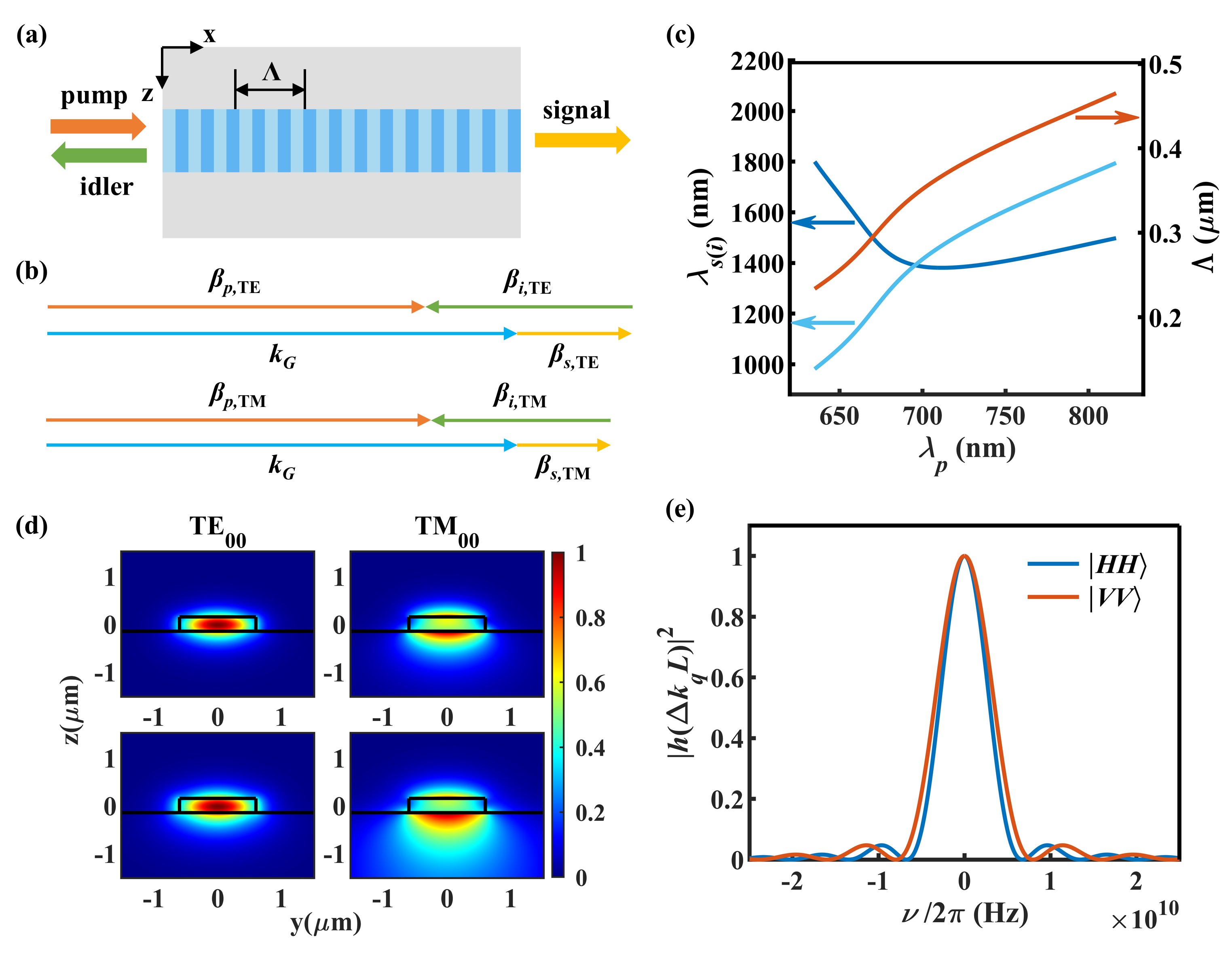}
		\caption{\label{fig6} The counter-propagating polarization-entangled photon pair generation scheme is illustrated in (a), and its QPM conditions are shown in (b). (c) Signal (idler) wavelength $\lambda_{s(i)}$ as a function of pump wavelength $\lambda_p$ that satisfies QPM condition (blue lines), and the corresponding poling periods $\Lambda$ (red line) for a waveguide with a thickness of 300 nm and a width of 1200 nm. (d) Mode profiles for fundamental TE/TM modes at $\lambda_s$ = 1439.26 nm (above) and $\lambda_i$ = 1679.20 nm (below), and (e) calculated bandwidth of these SPDC pairs.}
	\end{figure*}

	In addition to co-propagating polarization-entangled photon pairs, counter-propagating photon pair can also be generated based on type-0 scheme, as shown in Fig.~\ref{fig6}(a). In this configuration, the signal photon travels oppositely to idler photon. Consequently, the QPM condition changes to
	\begin{equation}\label{eq11}
		\beta_{p,\mathrm{TE}}-\beta_{s,\mathrm{TE}}+\beta_{i,\mathrm{TE}}-k_{G} = \beta_{p,\mathrm{TM}}-\beta_{s,\mathrm{TM}}+\beta_{i,\mathrm{TM}}-k_{G} = 0
	\end{equation}	
	as shown in Fig.~\ref{fig6}(b). Unlike the co-propagating scheme, a significantly larger $k_G$ is required to satisfy the QPM condition. Moreover, the range of waveguide parameters that fulfill the QPM condition is much narrower for the counter-propagating scheme. 
	
	Figure~\ref{fig6}(c) plots the relationship between the signal (idler) wavelength and the pump wavelength for a PPLN waveguide with a thickness of 300 nm and a width of 1200 nm. The SPDC pairs can be either degenerated at 1391 nm or non-degenerated with large detuning. When a pump light at 775 nm is lunched into the waveguide, polarization entangled counter-propagating photon pairs at 1439.26 nm and 1679.20 nm can be generated via SPDC, with a poling period of 429.81 nm. The corresponding mode profiles of the fundamental modes $\mathrm{TE_{00}}$ and $\mathrm{TM_{00}}$ for the signal and idler waves are shown in Fig.~\ref{fig6}(d). The bandwidths of the photon pairs are calculated and plotted in Fig.~\ref{fig6}(e), revealing that they are two orders of magnitude narrower (5.96 and 7.06 GHz) compared to the co-propagating pairs. While narrow bandwidths are desirable in many applications, they come at a cost. For the same pump power and waveguide length, the generation rate of entangled photon pairs is only $2.10\times10^5$ pairs/s. Additionally, the extremely small submicron-scale poling period poses significant challenges for sample fabrication. 
	
\section{Conclusion}
	In summary, we propose polarization-entangled photon pairs generation based on two type-0 SPDC in a thin-film PPLN waveguide with single poling period. The generated photon pairs can be either degenerate or non-degenerate with significant wavelength differences. While the co-propagating scheme provides high generation rates, the counter-propagating scheme achieves narrower bandwidths. By incorporating additional waveguide structures, temporal compensation can be achieved on the same chip. Our simple and flexible approach shows great potential in integrated quantum photonic applications.

\begin{acknowledgments}
	W.F. is partially supported by the National Natural Science Foundation of China (Grant No. 62035013, 62075192, 62475235), the Zhejiang Provincial Natural Science Foundation of China (Grant No. LZ23F050006) and the Quantum Joint Funds of the Natural Foundation of Shandong Province (No. ZR2020LLZ007).
\end{acknowledgments}

%\bibliography{Reference}
%

\end{document}